\documentclass{PoS}

\usepackage{tikz}
\usepackage{pgfplots}
\usetikzlibrary{positioning,arrows,patterns}
\usetikzlibrary{decorations.markings}
\usetikzlibrary{calc}
\usetikzlibrary{fit}
\usetikzlibrary{snakes}

\usepackage{amsmath}
\usepackage{amsthm}
\usepackage{bm}
\usepackage{mathtools}

\renewcommand{\vec}[1]{\bm{#1}}

\title{Electromagnetic corrections to leptonic decay rates of charged pseudoscalar mesons: finite-volume effects}

\ShortTitle{Electromagnetic corrections to leptonic decay rates}

\author{\speaker{N. Tantalo}\\
        Universit\`a di Roma ``Tor Vergata'' and INFN sezione di Roma ``Tor Vergata''\\
        E-mail: \email{nazario.tantalo@roma2.infn.it}}

\author{V. Lubicz\\
        Universit\`a di Roma Tre and INFN sezione di Roma Tre\\
        E-mail: \email{lubicz@fis.uniroma3.it}}

\author{G. Martinelli\\
        Universit\`a di Roma ``La Sapienza'' and INFN sezione di Roma\\
        E-mail: \email{guido.martinelli@roma1.infn.it}}

\author{C.T. Sachrajda\\
        Department of Physics and Astronomy, University of Southampton\\
        E-mail: \email{cts@soton.ac.uk}}

\author{F. Sanfilippo\\
        Department of Physics and Astronomy, University of Southampton\\
        E-mail: \email{f.sanfilippo@soton.ac.uk}}

\author{S. Simula\\
        INFN sezione di Roma Tre\\
        E-mail: \email{silvano.simula@roma3.infn.it}}

\abstract{
In ref.~\cite{Carrasco:2015xwa} we have recently proposed a method to calculate $O(e^2)$ electromagnetic corrections to leptonic decay widths of pseudoscalar mesons. The method is based on the observation that the infrared divergent contributions (that appear at intermediate stages of the calculation and that cancel in physical quantities thanks to the Bloch-Nordsieck mechanism) are universal, i.e. depend on the charge and the mass of the meson but not on its internal structure. In this talk we perform a detailed analysis of the finite-volume effects associated with our method. In particular we  show that also the leading $1/L$ finite-volume effects are universal and perform an analytical calculation of the finite-volume leptonic decay rate for a point-like meson.}

\FullConference{34th annual International Symposium on Lattice Field Theory\\
		24-30 July 2016\\
		University of Southampton, UK}

\begin{document}

\setlength{\parindent}{0pt}
\setlength{\parskip}{5pt}

\section{Introduction}
Electromagnetic radiative corrections cannot be neglected in the calculation of hadronic matrix elements when the target precision is at the percent level. For this reason in ref.~\cite{Carrasco:2015xwa} we have recently proposed a method to calculate $O(e^2)$ electromagnetic corrections to the leptonic decay widths of pseudoscalar mesons. 

\begin{figure}[!t]
\centering
\includegraphics[width=0.8\textwidth,]{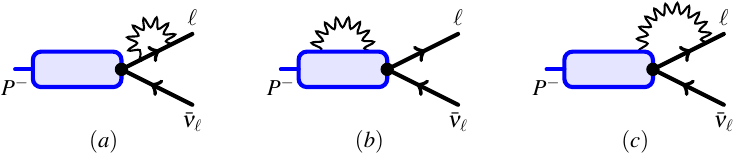}
\caption{\footnotesize Diagrams representing the $O(e^2)$ radiative corrections to $\Gamma_{V}(L)$.}
\label{fig:contributions}
\end{figure}

In our method we cope with the infrared divergences that appear at intermediate stages of the calculations by using the Bloch-Nordsieck approach. In this approach the $O(e^2)$ leptonic decay rate of a pseudoscalar meson is given by the sum of the so-called ``real'' and ``virtual'' contributions,
\begin{flalign}
\Gamma(E)/\Gamma^{\text{tree}} =
1 + e^2\lim_{\varepsilon\to 0}\left\{
\Gamma_V(\varepsilon) + \Gamma_R(\varepsilon,E)\right\}\;.
\label{eq:starting}
\end{flalign}
In the previous expression
\begin{flalign}
\Gamma^{\text{tree}}=\frac{G_F^2\, \vert V_{CKM}\vert^2}{8\pi}\,  f_P^2\, m_P^3 r_\ell^2\left(1-r_\ell^2\right)^2\;,
\qquad
r_\ell=\frac{m_\ell}{m_P}\;,
\label{eq:gammatree}
\end{flalign}
where $f_P$ is the leptonic decay constant of the meson in QCD, an unphysical quantity once radiative corrections are taken into account, while $m_P$ and $m_\ell$ are the physical masses (that include radiative corrections) of the meson and of the lepton. The real contribution $\Gamma_R(\varepsilon,E)$ is an inclusive quantity. It is obtained by including a soft photon in the final state and by integrating the exclusive three-body decay rate over the photon energy $E_\gamma$ in the range $E_\gamma\in[0,E]$. The virtual contribution $\Gamma_V(\varepsilon)$ has no real photons but a single virtual photon emitted and reabsorbed from the charged hadron and the charged lepton in all possible ways, see Figure~\ref{fig:contributions}. In fact the separation of the real and virtual contributions is unphysical and can only be done in presence of an infrared regulator, that we called $\varepsilon$ in the previous expressions. The infrared divergent contribution present in the real part, a term $c_{IR}\log(m_P^2/\varepsilon^2)$, is cancelled by the contribution $c_{IR}\log(\varepsilon^2/E^2)$ present in the virtual part.

In ref.~\cite{Carrasco:2015xwa} we built our method on the fact that the coefficient $c_{IR}$ is \emph{universal}, it depends on the electric charges and the masses of the particles but not on their internal structure. Therefore, by considering a theory in which the meson is treated as a point-like particle and by computing in this theory the virtual contribution $\Gamma_V^{pt}(\varepsilon)$, both the quantities
\begin{flalign}
\Gamma_{SD}(\varepsilon) = \Gamma_V(\varepsilon) - \Gamma_V^{pt}(\varepsilon)\;,
\qquad
\hat \Gamma(\varepsilon,E) = \Gamma_V^{pt}(\varepsilon) + \Gamma_R(\varepsilon,E) 
\end{flalign}
are infrared finite and can be computed separately, possibly with two different infrared regulators. Because of the quantization of spatial momenta on a finite volume it is very challenging to compute $\Gamma_R(\varepsilon,E)$ by using the volume itself as infrared regulator. For this reason we proposed to compute $\hat \Gamma(m_\gamma,E)$ directly in infinite volume by regulating the infrared with a photon mass $m_\gamma$ and by neglecting the structure-dependent contributions in $\Gamma_R(m_\gamma,E)$, terms of $O(E/\Lambda_{QCD})$. The systematics associated with this approximation have been quantified and, since this is not the subject of this talk, will not be discussed here. On the other hand, we proposed to calculate $\Gamma_{SD}(L)$ in a lattice QCD$+$QED simulation by using the volume $L$ as infrared regulator and to rewrite eq.~(\ref{eq:starting}) as
\begin{flalign}
\Gamma(E)/\Gamma^{\text{tree}} =
1
+ e^2\lim_{L\to \infty} \Gamma_{SD}(L)
+ e^2\lim_{m_\gamma\to 0} \hat \Gamma(m_\gamma,E)\;.
\end{flalign}

The main subject of this talk is $\Gamma_{SD}(L)$. In the following we shall show that
\begin{flalign}
&
\Delta \Gamma_{V}^{pt}(L)= 
c_{IR}\log(L^2 m_P^2) + \frac{c_1}{m_P L} + O\left(L^{-2}\right)\;,
\nonumber \\
&
\Delta \Gamma_{V}(L)= 
c_{IR}\log(L^2 m_P^2) + \frac{c_1}{m_P L} + O\left(L^{-2}\right)\;,
\label{eq:circ1}
\end{flalign}
where for any observable $\mathcal{O}$ we define the associated finite-volume effects as 
\begin{flalign}
\Delta \mathcal{O}(L) = \mathcal{O}(L)-\mathcal{O}(\infty)\;.
\end{flalign}
Eqs.~(\ref{eq:circ1}) mean that not only $c_{IR}$, but also the coefficient $c_1$ of the leading finite-volume effects on $\Gamma_{V}(L)$ is \emph{universal}, i.e. it is the same in the full theory and in the point-like approximation. We shall see that this is a consequence of electromagnetic gauge invariance and, in fact, we shall derive the very important result that the finite-volume effects on $\Gamma_{SD}(L)$ are $O(L^{-2})$. We shall also provide an analytical expression for $\Gamma_{V}^{pt}(L)$, a key ingredient for the implementation of our method in numerical simulations. 
%
\section{Universality}
In the introduction we defined universal those quantities that depend upon the masses and the electric charges of the particles but not on their internal structure. Then we also said that a quantity is universal if can be computed by using an effective theory of elementary charged particles. These two statements are in correspondence with the two different approaches that have been followed in the literature to prove the universality of the $1/L$ and $1/L^2$ terms of the finite-volume effects on the mass of charged hadrons. In the first approach one starts from the full theory (QCD$+$QED) and, by using skeleton expansion (see ref.~\cite{Borsanyi:2014jba}) or spectral analysis (see ref.~\cite{Lucini:2015hfa}) techniques, shows that universality is a consequence of the electromagnetic Ward identities of the theory. In the second approach, by resting on gauge invariance, one builds the effective lagrangian of a composite charged particle interacting with soft photons and gets the same result for the universal terms, see refs.~\cite{Davoudi:2014qua,Fodor:2015pna}. The two approaches are equivalent in a \emph{local} finite-volume formulation of QED. In this written version of the talk I shall discuss the universality of $c_{IR}$ and $c_1$ by using the effective theory approach. For a proof based on the first approach see Appendix~A of ref.~\cite{Lubicz:2016xro}.

In absence of electromagnetic interactions the following euclidean lagrangian
\begin{flalign} 
\mathcal{L}^{\text{tree}}=
\phi_P^\star(x)\left\{-\partial^2 + m_P^2\right\}\phi_P(x)+
\sqrt{2}G_F\, f_P\,\left\{ i\partial_\mu \phi_P(x)\, L^\mu(x) + \mbox{h.c} \right\}\;,
\label{eq:point0}
\end{flalign} 
is matched on the full theory in order to reproduce $\Gamma^{\text{tree}}$. In the previous expression $\phi_P$ is the field of the meson while $L^\mu=  \bar \ell \gamma^\mu \nu$ is the weak leptonic current and we have not shown the kinetic terms of the leptons. In order to understand how to construct the effective lagrangian with photons it is instructive to first discuss the range of applicability of $\mathcal{L}^{\text{tree}}$. This lagrangian can be used to calculate scattering amplitudes at $O(G_F)$ for processes in which the external states are electrically charged (possibly flavoured) and the momentum transfer is such that $(\sum_i p_i)^2< -(m_P+2m_\pi)^2$, where the sum runs over the (euclidean) momenta of the initial particles. Double insertions (loops) of the weak operator would lead to wrong results, thus signalling the absence in $\mathcal{L}^{\text{tree}}$ of operators that are relevant at higher orders in $G_F$. Moreover, notice that the amplitude for the scattering of a $P^+$ and a $P^-$ computed with $\mathcal{L}^{\text{tree}}$ gives zero. This is because there is no interaction of the field $\phi_P$ with itself or with any other hadron. In order to study processes in which other hadrons can be produced (even if these hadrons are not present in the final state) one has to add to $\mathcal{L}^{\text{tree}}$ the corresponding degrees of freedom, their self-interaction and the interaction with $\phi_P$. On the other hand, as far as on-shell quantities are concerned and in the range of validity of $\mathcal{L}^{\text{tree}}$, an operator such as $(\phi_P^\star \phi_P)^2$ has the only effect of renormalizing the fields and the parameters of $\mathcal{L}^{\text{tree}}$. This explains the restriction to low-energy processes in the charged sector of the theory and the absence in $\mathcal{L}^{\text{tree}}$ of hadronic interactions.

In the same range of validity of $\mathcal{L}^{\text{tree}}$ we can build an effective theory that describes the interaction of $P^\pm$ with soft photons. Also in this case the lagrangian will not contain hadronic interactions. At $O(G_F^0)$ we have to consider all the gauge invariant operators that are bilinear in the meson field. The problem is greatly simplified if we limit the validity of the theory to $O(e^2)$. Let us consider, for example, the following operator
\begin{flalign}
\phi_P^\star(x)D_{\mu}D_{\nu}D_{\mu}D_\nu\phi_P(x) \;,
\end{flalign}
where, since the meson has the same charge of the lepton,  $D_\mu=\partial_\mu -ieA_\mu(x)$. By using the commutation relation $[D_\mu, D_\nu] = -ieF_{\mu\nu}$, the equations of motions 
\begin{flalign}
\left\{-D^2 + m_P^2 \right\}\phi_P(x) =0\;,
\qquad
\partial_\mu F_{\mu\nu}(x) = -ie\left\{2 \phi_P^\star D_\nu\phi_P-\partial_\nu\left(\phi_P^\star \phi_P \right)\right\}(x) \;,
\end{flalign}
and by allowing for mass-dependent renormalizations, the previous operator reduces to a linear combination of the following operators
\begin{flalign}
\phi_P^\star\phi_P\, F_{\mu\nu}^2 \;,
\qquad 
\partial_\nu\left\{ \phi_P^\star\phi_P\partial_\mu F_{\mu\nu}\right\} \;,
\qquad
\partial_\nu\left\{F_{\nu\rho}\partial_\mu F_{\mu\rho}\right\}^2
\;,
\end{flalign}
two of which are total derivatives and do not contribute to the action. With similar arguments it is easy to show that a complete basis of operators for the $O(e^2,G_F^0)$ effective lagrangian is given by
\begin{flalign}
G^{1,n}=\phi_P^\star  \phi_P \partial^{2n}\left\{F_{\mu\nu}\right\}^2\;,
\qquad
G^{2,n}=\phi_P^\star\phi_P\partial^{2n}\left\{\partial_\mu F_{\mu\nu}\right\}^2\;,
\qquad
n=0,1,\cdots\;.
\label{eq:opeven}
\end{flalign}
In order to understand the weak sector of our effective lagrangian, up to and including $O(e^2 G_F)$, we need to consider all the gauge invariant operators linear in $\phi_P$ that can mix with the hadronic weak current $D_\mu \phi_P$ under strong radiative corrections. Operators proportional to $D^2 \phi_P$ can be eliminated with the equations of motions (and renormalization) and a complete basis is given by 
\begin{flalign}
&
W^{1,n}_\mu=\left\{ \partial^{2n} F_{\mu\nu}\right\} D_\nu \phi_P\;,
\qquad \ \
W^{2,n}_\mu=\left\{ \partial^{2n} \tilde F_{\mu\nu}\right\} D_\nu \phi_P\;,
\qquad \ \ \
W^{3,n}_\mu=\left\{ \partial^{2n} \partial_\nu F_{\nu\mu}\right\} \phi_P\;,
\nonumber \\
&
W^{4,n}_\mu=\left\{ \partial^{2n} F_{\alpha\beta}F_{\alpha\beta}\right\} D_\mu \phi_P\;,
\quad
W^{5,n}_\mu=\left\{ \partial^{2n} \tilde F_{\alpha\beta}F_{\alpha\beta}\right\} D_\mu \phi_P\;,
\quad
W^{6,n}_\mu=\left\{ \partial^{2n} F_{\mu\beta}\partial_\alpha F_{\alpha\beta}\right\} \phi_P\;,
\nonumber \\
&
W^{7,n}_\mu=\left\{ \partial^{2n} \tilde F_{\mu\beta}\partial_\alpha F_{\alpha\beta}\right\} \phi_P\;,
\quad \
W^{8,n}_\mu=\left\{ \partial^{2n} \partial_\alpha F_{\alpha\beta}\partial_\rho F_{\rho\beta}\right\} D_\mu \phi_P\;,
\end{flalign}
where $\tilde F_{\mu\nu}=\epsilon_{\mu\nu\alpha\beta}F_{\alpha\beta}$.
New operators appear also in the purely photonic sector giving rise to the well-known Euler-Heisenberg lagrangian. These operators do not contribute however at $O(e^2)$ when the external states are electrically charged. We stress again that at higher orders in $G_F$ additional operators become relevant. The new effective lagrangian is thus given by
\begin{flalign}
\mathcal{L}
=
\phi_P^\star\left\{-D^2 + m_P^2\right\}\phi_P
+
\sqrt{2}G_F f_P \left(L^\mu\, iD_\mu \phi_P +\mbox{h.c}\right)
+
\sum_{n=0}^\infty\left\{\sum_{i=1}^2 g^{i,n}\, G^{i,n}
+
L^\mu \sum_{i=1}^8w^{i,n}\, W^{i,n}_\mu + \mbox{h.c} \right\}.
\label{eq:point1}
\end{flalign}
The \emph{structure-dependent coefficients} are dimensional quantities that depend upon the
energy scales of the strong interactions and that, by neglecting exponentially suppressed finite-volume effects, can be chosen to be volume-independent: $g^{i,n} = g^{i,n}(m_\pi,\Lambda_{QCD})$ and $w^{i,n} = w^{i,n}(m_\pi,\Lambda_{QCD})$.

We are now going to explain how universality follows from the structure of $\mathcal{L}$. At $O(e^2)$, the finite-volume effects on a generic observable can be written as
\begin{flalign}
\Delta \mathcal{O}(L) = \hat \sum_{\vec k}\int \frac{dk_0}{2\pi} 
\frac{I_{\mathcal{O}}(k,p_i)}{k^2}\;,
\qquad
\hat \sum_{\vec k}= \frac{1}{L^3}\sum_{\vec k\in \Omega/L}-\int \frac{d^3k}{(2\pi)^3}\;,
\label{eq:zeros}
\end{flalign}
where $k$ is the momentum of the (single) virtual photon, $p_i$ are external momenta, $I_{\mathcal{O}}(k,p_i)/k^2$ is the loop integrand corresponding to the observable $\mathcal{O}$ and we have shown explicitly the factor $1/k^2$ coming from the photon propagator. The sum over $\vec k$ runs over the set $\Omega/L$ of spatial momenta allowed by the boundary conditions and the integral over $d^3k$ is the infinite volume result. Power-law finite-volume effects arise if the function resulting from the $dk^0$ integral in eqs.~(\ref{eq:zeros}) is not infinitely differentiable for real values of $\vec k$ (the difference between the integral of a regular function and its approximation as a Riemann sum vanishes faster than any power). In a local formulation of the theory this happens only at the infrared singular point $k^2=0$. In this case, by studying the behaviour of the integrand around the singularity one gets the following simple rule
\begin{flalign}
I_{\mathcal{O}}(k,p_i)\, k^{-2} \sim z_{\mathcal{O}}(\hat k,p_i)\, k^{-\alpha_{\mathcal{O}}}\;,
\qquad
\Delta \mathcal{O}(L) = O\left(L^{\alpha_{\mathcal{O}}-4}\right)\;,
\label{eq:rule}
\end{flalign}
where $\hat k_\mu= k_\mu/\sqrt{k^2}$. For $\alpha_{\mathcal{O}}=4$ one has a logarithmic infrared divergence. In the calculation of $\Delta \Gamma_{V}(L)$ the infrared divergent contributions come from the universal part of $\mathcal{L}$ and are given by
\begin{flalign}
\Delta \Gamma_{V}(L) \supset 
\hat \sum_{\vec k}\int \frac{dk_0}{2\pi} 
\frac{1}{k^2}\left\{
-\frac{4m_P^2}{\left(2p_P\cdot k + k^2\right)^2}
+
\frac{8p_P\cdot p_\ell}{\left(2p_P\cdot k + k^2\right)(2p_\ell\cdot k + k^2)}
\right\}\;,
\end{flalign}
where $p_P^2=-m_P^2$ is the on-shell momentum of the meson, $p_\ell^2=-m_\ell^2$ that of the lepton and $A\supset B$ means that $B$ is a term contributing to $A$. The first term in curly brackets comes from the wave-function renormalization of the meson, a diagram of class $(b)$ in Figure~\ref{fig:contributions}, while the second comes from the vertex correction, a diagram of class $(c)$ in Figure~\ref{fig:contributions}. Notice that the contributions of the diagrams of class $(a)$ in Figure~\ref{fig:contributions}, that we call $\Gamma_{\ell\ell}(L)$, are the same in the full theory and in the point-like approximation and cancel exactly in $\Gamma_{SD}(L)$. For this reason we do not need the finite-volume expression of $\Gamma_{\ell\ell}(L)$ in our calculation.

The leading structure-dependent, and therefore non-universal, contributions to $\Delta \Gamma_{SD}(L)$ are generated from insertions of the operators $W^{1,0}_\mu$ and $W^{2,0}_\mu$. These operators correspond to weak vertices with the emission of one or two photons that, because of the presence of the field tensor $F_{\mu\nu}$, are proportional to the photon momentum $k$. In the case of the emission of a single photon, this has to be attached to one of the charged external lines and we have
\begin{flalign}
\Delta \Gamma_{SD}(L) =
(w^{1,0}+w^{2,0})\hat \sum_{\vec k}\int \frac{dk_0}{2\pi}\, 
\left\{
\frac{O(k)}{k^2\left(2p_P\cdot k+k^2\right) } 
+
\frac{O(k)}{k^2\left(2p_\ell\cdot k+k^2\right) } 
\right\}
= O(L^{-2})\;.
\end{flalign}
The fact that the other structure-dependent operators present in $\mathcal{L}$ give sub-leading  contributions in $1/L$ can be understood with a similar analysis. To this end one has to notice that operators containing $\partial_\mu F_{\mu\nu}$ or $F_{\mu\nu}^2$ are proportional to two powers of the photon momentum $k$. Moreover, at $O(e^2)$ the vertices with two photons generate ``tadpole'' graphs that do not have the $1/p_i\cdot k$ factor coming from the charged particle propagators. This implies that the operators $G^{1,n}$ and $G^{2,n}$ contribute to the radiative corrections to the mass of the meson, see eq.~(\ref{eq:pretty}) below, but not to its wave-function renormalization. We have thus shown our main result, i.e. the universality of $c_{IR}$ and $c_1$ in eqs.~(\ref{eq:circ1}) and the fact that $\Delta \Gamma_{SD}(L)=O(L^{-2})$.

Before closing our discussion on universality we need to clarify the issue of locality. In ref.~\cite{Lucini:2015hfa} it has been shown that a local formulation of the finite-volume theory, called QED$_\text{C}$, can be obtained by enforcing charge-conjugation boundary conditions along the spatial directions. 
In this setup the photon is anti-periodic in space and the sum in eqs.~(\ref{eq:zeros}) is well defined. Structure-dependent finite-volume effects on the masses of charged hadrons are $O(L^{-4})$ in QED$_\text{C}$. This has been shown by using spectral decomposition techniques but, in fact, the same result can be readily obtained by using the effective lagrangian $\mathcal{L}$: the operators $G^{1,n}$ and $G^{2,n}$ give 
\begin{flalign}
\Delta m_P(L) \supset
\hat \sum_{\vec k}\int \frac{dk_0}{2\pi}\, 
\frac{g^{1,n}\, O(k^{2+2n})+g^{2,n}\, O(k^{4+2n})}{k^2} = O(L^{-4-2n})\;.
\label{eq:pretty}
\end{flalign}
With periodic boundary conditions in space the sum in eqs.~(\ref{eq:zeros}) is not well defined because $\vec k=\vec 0$ is contained in $\Omega/L$. A non-local formulation of the theory can be obtained in this case by considering the so-called QED$_\text{L}$ prescription~\cite{Hayakawa:2008an}, i.e. by quenching the spatial zero-modes of the photon field. In this theory additional power-law finite-volume effects arise. The degrees of freedom associated with the propagation of massive hadronic states generate $O(L^{-3})$ corrections in eqs.~(\ref{eq:zeros}) at non-vanishing values of $k^2$. For example, by taking the external meson at rest, $p_P=(im_P,\vec 0)$, and by assuming that a particle (multi-particle state) of  mass (rest energy) $m_h$ can propagate between the insertions of two electromagnetic currents, one has contributions to $\Delta m_P(L)$ of the form 
\begin{flalign}
\hat \sum_{\vec k}\int \frac{dk_0}{2\pi} \frac{\rho(k_0,\vec k)}{k^2\left\{(p_P+k)^2 + m_h^2 \right\}}
&\supset
-\left\{\frac{1}{L^3}\sum_{\vec k\in \Omega^\prime/L}-\int \frac{d^3k}{(2\pi)^3}\right\} 
\frac{\rho(-im_P-im_h,\vec k) + O(\vec k^2)}{2(m_P+m_h)^2\sqrt{\vec k^2+m_h^2}} 
\nonumber \\
&= \frac{\rho(-im_P-im_h,\vec 0)}{2m_h(m_P+m_h)^2 L^3} + O\left( e^{-\lambda L}\right)\;,
\end{flalign}
where $\Omega^\prime=2\pi\mathbb{Z}^3-\{\vec 0\}$ and $\lambda = O(m_P)$. 
These effects cannot be fully reproduced by an effective theory from which these states have been integrated out. 
In our case we have explicitly verified, by using skeleton expansion and spectral decomposition techniques, that the result $\Delta \Gamma_{SD}(L)=O(L^{-2})$ holds in QED$_\text{L}$, see ref.~\cite{Lubicz:2016xro}.

\section{Analytical computation of $\Gamma_{V}^{pt}(L)$}
In order to compute $\Gamma_{V}^{pt}(L)$ one has to start from the lagrangian $\mathcal{L}$ without structure-dependent operators, i.e. with $g^{i,n}=w^{i,n}=0$. The perturbative calculation, that we did in QED$_\text{L}$, is technically involved and cannot be described here in full details because of the page limits of this contribution\footnote{in the \texttt{arXiv} version of this contribution I have already taken two additional pages!}. For this reason we only quote our final result and explain how to use it. Our result is
\begin{flalign}
\Gamma^{pt}_V(L)-\Gamma_{\ell\ell}(L)
=
c_{IR} \log(L^2 m_P^2) + c_0 
+ \frac{c_1}{(m_P L)}
+ \frac{c_2}{(m_P L)^2}
+ \frac{c_3}{(m_P L)^3}
+ O(e^{-\lambda L})\;,
\end{flalign}
where $\lambda >0$ and\footnote{In the original version of this manuscript there was a typo in the coefficient of the $1/L^3$ term in eq.~(\ref{eq:nasty}) and, consequently, the coefficient $c_3$ in eq.~(\ref{eq:coeffs}) was not correct. The typo has been discovered from the authors of ref.~\cite{DiCarlo:2021apt} by performing an independent calculation.}
\begin{flalign}
&
c_{IR} = \frac{1}{8\pi^2}\left\{\frac{(1+r_\ell^2)\log(r_\ell^2)}{(1-r_\ell^2)} + 1\right\}\;,
\nonumber \\
&
c_0 
=
\frac{1}{16\pi^2}
\left\{
2\log\left(\frac{m_P^2}{m_W^2} \right)
+
\frac{
(2-6r_\ell^2)\log(r_\ell^2)
+
(1+r_\ell^2)\log^2(r_\ell^2)
}{1-r_\ell^2}
-\frac{5}{2}
\right\}
+ \frac{\zeta_C(\vec 0)-2\zeta_C(\vec \beta_\ell)}{2}\;,
\nonumber \\
&
c_1
=
-\frac{2(1+r_\ell^2)\zeta_B(\vec 0)}{1-r_\ell^2}
+\frac{8r_\ell^2\zeta_B(\vec \beta_\ell)}{1-r_\ell^4}\;,
\quad
c_2
=
\frac{4\zeta_A(\vec 0)}{1-r_\ell^2} 
-\frac{8\zeta_B^{P\ell}(\vec \beta_\ell)}{1-r_\ell^4}\;,
\quad
c_3
=
-\frac{4(2+r_\ell^2)}{\left(1+r_\ell^2\right)^3}\;.
\label{eq:coeffs}
\end{flalign}
The generalized $\zeta$-functions appearing in the previous expressions are dimensionless functions of the kinematical variable
$\vec \beta_\ell = \vec p_\ell/E_\ell = \vec{\hat p_\ell}(1-r_\ell^2)/(1+r_\ell^2)$, where $\vec p_\ell$ is the spatial momentum of the lepton when $p_P=(im_P,\vec 0)$. These arise in the calculation of finite-volume one-loop master integrals and can be computed with arbitrary numerical precision. For example,
\begin{flalign}
\frac{1}{L^3}\sum_{\vec k\in \Omega^\prime/L}\int\frac{dk^0}{2\pi}\, \frac{m_W^2}{
k^2(k^2+m_W^2)\,\left[2p_P\cdot k + k^2\right]}
=
\frac{1}{16\pi^2}\left\{1-\log\left(\frac{m_P^2}{m_W^2}\right)\right\}+
\frac{\zeta_B(\vec 0)}{(m_P L)} 
+ \frac{1}{8(m_P L)^3}\;,
\end{flalign}
where $\Omega^\prime=2\pi\mathbb{Z}^3-\{\vec 0\}$. Notice that we have regulated ultraviolet divergences in the so-called $W$-regularization scheme, see ref.~\cite{Carrasco:2015xwa} for more details, and we have neglected terms of $O(e^{-\lambda L})$ and of $O((m_W L)^{-3})$. The last are very peculiar of QED$_\text{L}$. The calculation of the $\zeta$-functions at $\vec \beta_\ell=\vec 0$ is rather standard and we quote below the numerical results needed in order to use eqs.~(\ref{eq:coeffs}),
\begin{flalign}
&
\zeta_A(\vec 0) = -0.22578495944(1)\;,\qquad 
\zeta_B(\vec 0) = -0.05644623986(1)\;,
\nonumber \\
&
\zeta_C(\vec 0) = -0.06215473226(1)\;.
\end{flalign}
On the other hand, the evaluation of finite-volume one-loop master integrals at $\vec \beta_\ell\neq\vec 0$ is a cumbersome exercise in numerical analysis. The trickiest master integral appearing in our calculation is the infrared divergent one, 
\begin{flalign}
&
-\frac{8p_P\cdot p_\ell}{L^3}\sum_{\vec k\in \Omega^\prime/L}\int\frac{dk^0}{2\pi}\, \frac{1}{
k^2\,
\left[2p_P\cdot k + k^2\right]\,
\left[2p_\ell\cdot k + k^2\right]
}  
\nonumber \\
&=
-\frac{(1+r_\ell^2)\log(r_\ell^2)}{16\pi^2(1-r_\ell^2)}
\left\{
2\log\left(L^2 m_P^2\right)
+
\log(r_\ell^2)
\right\} + \zeta_C(\vec{\beta_\ell})
+
\frac{(3+r_\ell^2)(5+2r_\ell^2+r_\ell^4)}{4(1+r_\ell^2)^3\, (m_P L)^3}\;.
\label{eq:nasty}
\end{flalign}
The following expression (see ref.~\cite{Lubicz:2016xro} for other representations)
\begin{flalign}
\zeta_C(\vec{\beta_\ell})
&=
\frac{1}{2\beta_\ell}\log\left(\frac{1+\beta_\ell}{1-\beta_\ell}\right)\,
\frac{\log(u_\star) + \gamma_E}{4\pi^2}
-\frac{4u_\star^{3/2}}{3\sqrt{\pi}}
\nonumber \\
&+
\frac{2}{\sqrt{\pi}}
\sum_{\vec k \in \Omega^\prime}
\frac{
\Gamma\left(\frac{3}{2},u_\star \vec k^2 \right)
}{\vert \vec k\vert^3\, \left[1- (\vec{\hat k}\cdot \vec \beta_\ell)^2\right]}
\left\{
1
+
\frac{
e^{u_\star(\vec{k}\cdot \vec \beta_\ell)^2}\,
\bar \Gamma\left[\frac{3}{2},u_\star (\vec{k}\cdot \vec \beta_\ell)^2 \right]
}{\vert\vec{\hat k}\cdot \vec \beta_\ell\vert\, e^{u_\star\vec k^2}\,\Gamma\left(\frac{3}{2},u_\star   
\vec k^2 \right)}
\right\}
\nonumber \\
&+
\frac{1}{4\pi^2}
\sum_{\vec n \neq \vec 0}
\int_0^{\frac{4u_\star}{\vec n^2}} \frac{du}{u}
e^{-\frac{1}{u}}\,
\int_0^{\frac{1}{1+\beta_\ell}}dy\, 
\frac{1-\frac{2y(\vec{\hat n} \cdot \vec{\beta_\ell})}{\sqrt{u(1-2\beta_\ell y)}}\, 
\mbox{D}  \left(\frac{y(\vec{\hat n} \cdot \vec{\beta_\ell})}{\sqrt{u(1-2\beta_\ell y)}}\right)}
{(1-2\beta_\ell y)}\;,
\end{flalign}
where $u_\star>0$ is an arbitrary parameter ($\zeta_C$ does not depend upon $u_\star$), $\gamma_E$ is the Euler-Mascheroni constant and
\begin{flalign}
&
\Gamma(\alpha,x) = \int_{x}^\infty du\, u^{\alpha -1}\, e^{-u}\;,
\quad
\bar \Gamma(\alpha,x) = \int_0^{x} du\, u^{\alpha -1}\, e^{-u}\;,
\quad  
\mbox{D}(x) = e^{-x^2}\int_0^x du\, e^{u^2}\;,
\end{flalign}
can be used to evaluate $\zeta_C(\vec{\beta_\ell})$ with arbitrary numerical precision. For example, at the physical values\footnote{We have used
$m_{\pi^+}=139.57018$~MeV, $m_{K^+}=493.677$~MeV and $m_\mu=105.65837$~MeV.} of the pion, kaon and muon masses we have
\begin{flalign}
\zeta_C(\vec{\beta_{\mu}^\pi})=-0.06331584128(1),\quad
\zeta_C(\vec{\beta_{\mu}^K})=-0.09037019089(1),\quad
\vec{\hat p_\mu} = \frac{(1,1,1)}{\sqrt{3}}\;.
\end{flalign}
The last two quantities needed in order to use eqs.~(\ref{eq:coeffs}) are
\begin{flalign}
\zeta_B^{P\ell}(\vec{\beta_\ell})
=&
-
\left\{
\sqrt{\frac{u_\star}{\pi}}
+
\frac{1}{16\pi^2 u_\star(1-\beta_\ell^2)}
\right\}
+
\sum_{\vec k\in \Omega^\prime}\,
\frac{
\Gamma\left(\frac{1}{2},u_\star \vec k^2 \right)
\left\{
1+
\vert\vec{\hat k}\cdot\vec{\beta_\ell}\vert\,  
\frac{
e^{u_\star(\vec{k}\cdot\vec{\beta_\ell})^2}
\bar \Gamma\left(\frac{1}{2},u_\star(\vec{k}\cdot\vec{\beta_\ell})^2 \right)}
{
e^{u_\star\vec k^2}\,
\Gamma\left(\frac{1}{2},u_\star \vec k^2 \right)}\right\}
}
{2\sqrt{\pi}\vert \vec k\vert\left\{ 1-(\vec{\hat k}\cdot\vec{\beta_\ell})^2\right\}}
\nonumber \\
&+
\frac{1}{4\pi^2(1-\beta_\ell^2)}
\sum_{\vec n \neq \vec 0}
\frac{1}{\vec n^2}
\int_0^{\frac{4u_\star}{\vec n^2}}\frac{du}{u^2}\, e^{-\frac{1}{u}} 
\left\{
1-
\frac{2\vec{\hat n}\cdot\vec{\beta_\ell}}{\sqrt{u(1-\beta_\ell^2)}}
\mbox{D}\left(\frac{\vec{\hat n}\cdot\vec{\beta_\ell}}{\sqrt{u(1-\beta_\ell^2)}}\right)
\right\}\;,
\nonumber \\
\nonumber \\
\zeta_B(\vec \beta_\ell)
=&
-\left(
\frac{u_\star}{4}
+
\frac{1}{16\pi^{\frac{3}{2}}\, \sqrt{u_\star(1-\beta_\ell^2)}}
\right)
+
\sum_{\vec k\in \Omega^\prime}
\frac{
\Gamma\left(1,u_\star\vec k^2\, \left[1 - (\vec{\hat k} \cdot \vec{\beta_\ell})^2\right] \right)
}
{4\vec k^2\, \left[1 - (\vec{\hat k} \cdot \vec{\beta_\ell})^2\right]}
\nonumber \\
&+
\frac{1}{16\pi^{\frac{3}{2}}\sqrt{1-\beta_\ell^2}}
\sum_{\vec n \neq \vec 0}
\frac{
\Gamma\left(\frac{1}{2},\frac{\vec n^2}{4u_\star}\, 
\left[1+\frac{\left(\vec {\hat n}\cdot \vec {\beta_\ell}\right)^2}{1-\beta_\ell^2}\right]\right)
}{\vert \vec n\vert\, \sqrt{1+\frac{\left(\vec {\hat n}\cdot \vec {\beta_\ell}\right)^2}{1-\beta_\ell^2}}}\;.
\end{flalign}
At the physical values of the pion, kaon and muon masses we have
\begin{flalign}
&
\zeta_B^{P\ell}(\vec{\beta_{\mu}^\pi})=-0.23173738346(1),\quad
\zeta_B^{P\ell}(\vec{\beta_{\mu}^K})=-0.45599283983(1),\quad
\vec{\hat p_\mu} = \frac{(1,1,1)}{\sqrt{3}}\;,
\nonumber \\
&
\zeta_B(\vec{\beta_{\mu}^\pi})=-0.05791071589(1),\quad \ \ 
\zeta_B(\vec{\beta_{\mu}^K})=-0.10350847338(1).
\end{flalign}
%

\section{Conclusions}
We have performed a detailed theoretical analysis of the finite-volume effects associated with the method of ref.~\cite{Carrasco:2015xwa} 
for the evaluation of $O(e^2)$ radiative corrections to leptonic decay widths of pseudoscalar mesons.
We have shown that the coefficients of the infrared logarithm and of the leading $O(L^{-1})$ finite-volume effects are universal, i.e. do not depend on the internal structure of the meson. Universal quantities can be computed analytically in a theory of elementary charged particles and we performed such a calculation. The analytical formulae presented here can be used to correct the results of lattice simulations. The finite-volume effects that remain after the analytical correction are structure-dependent and of $O(L^{-2})$. 
All the ingredients are now in place for the application of our method in lattice simulations and, in fact, preliminary numerical results for the $\pi^-\to \mu\bar \nu_\mu(\gamma)$ and $K^-\to \mu\bar \nu_\mu(\gamma)$ decay rates have been presented at this conference, see ref.~\cite{Lubicz:2016mpj}. An extended discussion of the results presented here can be found in ref.~\cite{Lubicz:2016xro}.

\acknowledgments{N.T. warmly thanks A.~Patella, for a critical reading of the manuscript and for illuminating discussions on the subjects covered in this talk, and the authors of ref.~\cite{DiCarlo:2021apt} for pointing out a typo in eq.~(\ref{eq:nasty}).}



{\footnotesize
\begin{thebibliography}{99}
\bibitem{Carrasco:2015xwa}
  N.~Carrasco {\it et al.},
  Phys.\ Rev.\ D {\bf 91} (2015) no.7,  074506
  [arXiv:1502.00257 [hep-lat]].

\bibitem{Borsanyi:2014jba}
  S.~Borsanyi {\it et al.},
  Science {\bf 347} (2015) 1452
  [arXiv:1406.4088 [hep-lat]].

\bibitem{Lucini:2015hfa}
  B.~Lucini, A.~Patella, A.~Ramos and N.~Tantalo,
  JHEP {\bf 1602} (2016) 076
  [arXiv:1509.01636 [hep-th]].

\bibitem{Davoudi:2014qua}
  Z.~Davoudi and M.~J.~Savage,
  Phys.\ Rev.\ D {\bf 90} (2014) no.5,  054503
  [arXiv:1402.6741 [hep-lat]].

\bibitem{Fodor:2015pna}
  Z.~Fodor {\it et al.},
  Phys.\ Lett.\ B {\bf 755} (2016) 245
  [arXiv:1502.06921 [hep-lat]].

\bibitem{Lubicz:2016xro}
  V.~Lubicz {\it et al.},
  arXiv:1611.08497 [hep-lat].
  
\bibitem{Hayakawa:2008an}
  M.~Hayakawa and S.~Uno,
  Prog.\ Theor.\ Phys.\  {\bf 120} (2008) 413
  [arXiv:0804.2044 [hep-ph]].

\bibitem{Lubicz:2016mpj}
  V.~Lubicz {\it et al.},
  arXiv:1610.09668 [hep-lat].

\bibitem{DiCarlo:2021apt}
M.~Di Carlo, M.~T.~Hansen, N.~Hermansson-Truedsson and A.~Portelli,
[arXiv:2109.05002 [hep-lat]].

\end{thebibliography}
}
\end{document}